# Complete Structural Determination of Mesostructural Dodecagonal Quasicrystalline Particles

Xueliang Zhang,[1,+] Xi Wang,[1,+] Nobuhisa Fujita,[2] Osamu Terasaki[3] and Lu Han[1,*]

[1]School of Chemical Science and Engineering, Tongji University, 1239 Siping Road, Shanghai, 200092, China.

[2]Institute of Multidisciplinary Research for Advanced Materials, Tohoku University, Sendai 980-8577, Japan.

[3]School of Physical Science and Technology, Centre for High-resolution Electron Microscopy and Shanghai Key Laboratory of High-resolution Electron Microscopy, ShanghaiTech University, 201210, Shanghai, China.

[+]These authors contributed equally to this work.

*e-mail: luhan@tongji.edu.cn

## Abstract

Quasicrystals have revolutionized our understanding of order in solids[1,2] by demonstrating exotic structural and physicochemical properties with diverse potential applications.[3-6] Despite the development of various theoretical models and experimental techniques to describe quasicrystal structures,[7-9] the precise determination of local three-dimensional (3D) arrangements of constituent atoms, or of secondary building units such as clusters or micelles, remains elusive. This challenge is particularly acute in self-assembled soft-matter quasicrystalline systems, where the complex assembly of molecular groups introduces additional defects and structural modulations.[10] Herein, we report the first complete structural determination of self-assembled mesostructural dodecagonal quasicrystalline particles. Employing advanced electron tomography, combined with dedicated structural tracing and processing workflows, the 3D coordinates of all nodal sites were extracted. This approach reveals that the actual structure deviates from the conventionally assumed tetrahedral close packing geometry, exhibiting diverse coordination environments and displacive fluctuations. We identified and quantified rotational intergrowths arising from node exchange, as well as various defects and disorder, with these features discernible only through 3D analysis. Additionally, we propose a simplified two-layer stacking of isomorphic hexagonal model to form dodecagonal quasicrystal. This work advances our understanding of soft-matter dodecagonal quasicrystals and paves the way for detailed structural elucidation of self-assembled systems.

## Introduction

Accurate structure determination is essential for understanding the physicochemicl properties and formation mechanisms of materials. However, the structural details of quasicrystals and their formation processes remain far from fully understood.[11] Commonly used diffraction techniques, such as X-ray diffraction,[12,13] electron diffraction[1,5,14] and neutron diffraction,[15] generally provide average information in reciprocal space. While decorated tiling models fitted to diffraction data within a higher-dimension framework have been accepted as a reasonable approximation to the atomic structure of quasicrystals,[16,17] direct real-space determintion of the structure remains highly desirable to eliminate existing ambiguities in structural understanding. Meanwhile, (scanning) transmission electron microscopy ((S)TEM) have been widely used to visualize projected tiling patterns and extract local structural information.[5,8,10] However, the inherent loss of information in two-dimensional (2D) projections often obscures the three-dimensional (3D) structure, making it difficult to retrieve.

As an important category of quasicrystal, dodecagonal quasicrystals have been discovered not only in metallic alloys[18,19] but also, more recently, in soft-matter systems, such as dendritic polymer liquid crystals,[20] ABC star-shaped polymers,[21,22] colloidal nanoparticle superstructures,[23,24] silica mesoporous crystals,[10,25] etc. In addition to their promise for developing next-generation photonics and metamaterials, these materials offer unique opportunities to investigate quasicrystal structures from a perspective that beyond the atomic scale. These soft-matter quasicrystals often display geometric motifs and symmetries resembling those of atomic quasicrystals. Yet their molecular building units, with mesoscale dimensions, introduce diverse packing behaviors and flexibility, thereby allowing new structural possibilities and modulations of the lattice or basis through inherently soft assembly processes. Consequently, deciphering the quasiperiodic order in soft-matter quasicrystals is even more challenging. To date, a complete structural solution that determines the exact positions of all structural units in quasicrystals remains elusive, posing a significant obstacle to deeper insight into their structures, properties, and underlying mechanisms.

Herein, we present a method for the complete determination of mesostructures in dodecagonal quasicrystalline particles formed through the cooperative self-assembly of cationic surfactant and silica precursor mediated by the coupling effect of organosilane. Our approach integrates electron tomography with advanced reconstruction algorithms to generate high-quality 3D tomograms. This enables us to extract the positions of individual nodes (i.e., cavities which, during synthesis, used to be occupied by surfactant micelles enclosed within slica) using a custom tracing procedure. This workflow facilitates full 3D analysis of the quasicrystal structure, revealing the microscopic origins of quasiperiodicity and enabling detailed examination of local structural features such as grain boundaries, positional fluctuations, and defects.

## Results

The selected area electron diffraction (SAED) patterns of randomly chosen particles in the sample all exhibit 12-fold rotational arrangements of the diffraction spots, indicating a non-crystallographic, dodecagonal order (Supplementary Figure S1). Figure 1a shows a bright-field TEM image of a representative particle (QC-1), viewed along its primary 12-fold rotational axis (taken as the $z$-axis). The corresponding fast Fourier transform (FFT) also confirms the 12-fold dodecagonal symmetry (inset). The brighter contrast, corresponding to projected regions of low electrostatic potential, represents nodes (or cavities) encaged by the silica framework. The brightest spots in the image were identified as the vertices of a tiling pattern composed of equilateral triangles and squares, as shown in Figure 1a. The ratio of triangles to squares is $92/41 \approx 2.24$, which is close to the ideal ratio $4/\sqrt{3} \approx 2.31$ of the dodecagonal square-triangle tiling.[26] Additionally, phason-strain analysis was carried out to quantify the deviation of the observed structure from ideal dodecahedral order. The results indicate that the two phason parameters ($\lambda$ and $d$, which quantify the linear phason strain and random phason disorder, respectively) are relatively small, indicating that the dodecagonal quasiperiodic order is well preserved (Supplementary Table S1).

To fully resolve the 3D architecture of QC-1, we employed electron tomography using 57 tilt-series bright-field TEM images collected over a goniometer range of -52.42° to +60.06° with ~2° increments to mitigate sample damage (Figure 1b, all images and angles are listed in Supplementary Figure S2). After contrast transfer function (CTF) correction, the image stack was aligned to compensate for drift and tilt-axis deviation. The 3D volumetric data were then reconstructed using the REal Space Iterative Reconstruction (RESIRE) algorithm.[27] The resulting tomogram (Figure 1c, Supplementary Movie S1) exhibits a slab-shaped morphology and displays structural features consistent with the contrast observed along the 12-fold rotational $z$-axis in Figure 1a. In the tomogram, brighter contrast signifies higher electron-beam intensities that correspond to lower material density, thereby emphasizing the nodes (cavities).

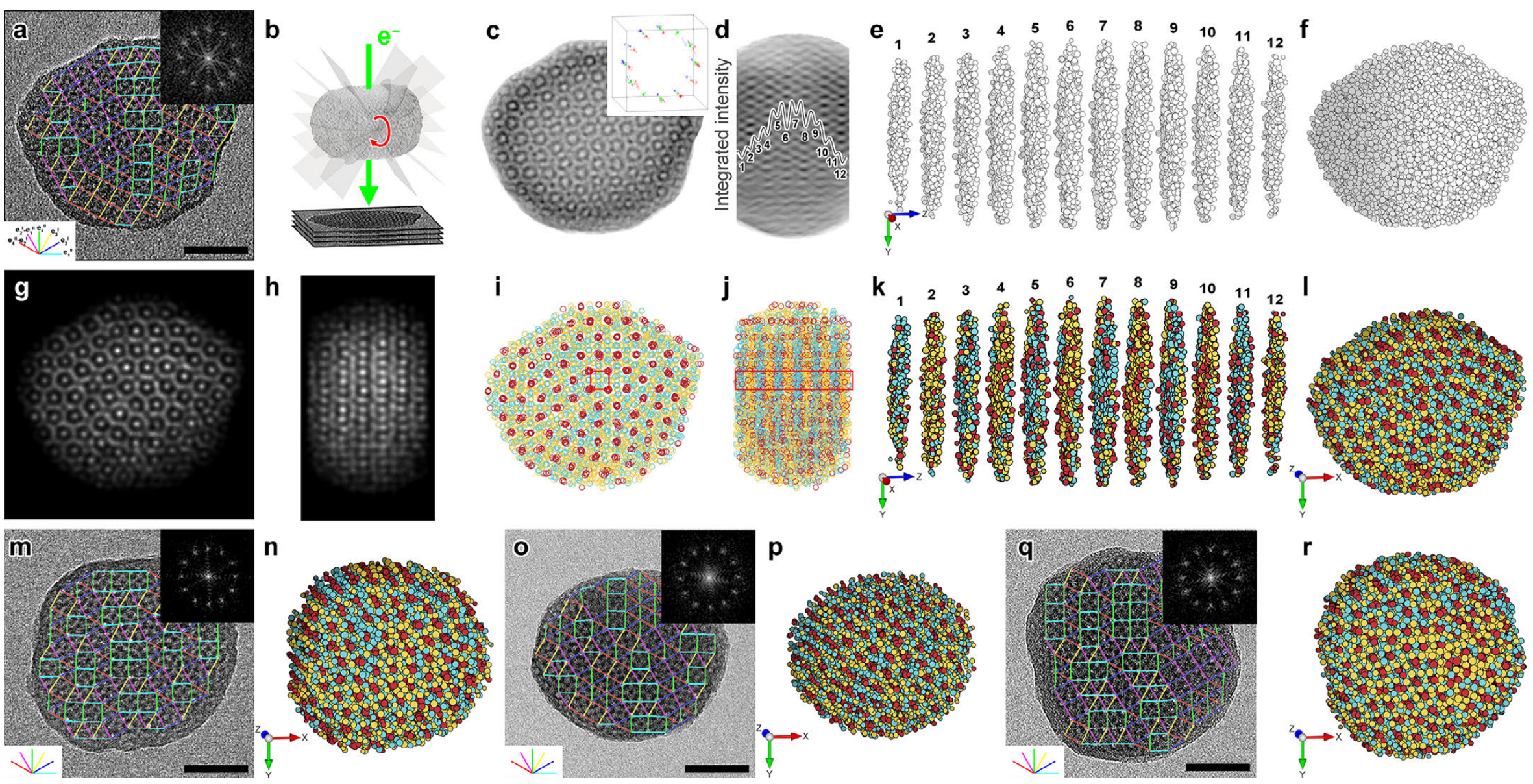

**Figure 1.** 3D determination of node coordinates in mesostructural dodecagonal quasicrystal particles. (a) 2D TEM image of a representative particle (QC-1), overlaid with 2D square-triangle tiling pattern; the inset shows the corresponding FFT with 12-fold rotational symmetry. (b) Schematic drawing of the data acquisition using tilt-series TEM images for 3D tomographic reconstruction. (c) Reconstructed 3D tomogram projected along the 12-fold axis (*z*-axis) along with its 3D FFT (inset), exhibiting 12-fold symmetry. Observed peaks in the FFT are colored according to whether can be indexed as $hkml\bar{1}$, *hkml*0 and *hkml*1, with *h, k, m,* and *l* representing the four Miller indices along the dodecagonal quasiperiodic plane. (d) Side projection of the tomogram overlaid with the integrated intensity profile. The peaks labeled 1 to 12 are indicative of the layered structure within the particle. (e) 12 slices obtained based on the peak positions shown in (d). The extracted 2D coordinates of the nodes within each slice were subsequently lifted to 3D coordinates. (f) Reconstructed 3D coordinates of nodes within the particle. (g) Simulated TEM image of the particle taken from the front projection. The simulation was performed by excising nodes from a solid body, with regions outside the particle rendered in black. However, the simulation remains a bright-field image, in which bright contrast corresponds to regions of low electrostatic potential within the particle. (h) TEM simulation of the particle from side projection. (i) Categorization of experimental nodal positions. The vertices of all equilateral triangles and squares were designated as Type I (red), while the surrounding nodes were alternately labeled as Type II (yellow) and Type III (cyan). (j) Side projection of the categorized nodes, consistent with the layers identified in (d). (k) Lateral layered structure expanded along the *z*-axis, showing the alternating arrangement of nodal types II and III. (l) 3D model of the whole particle, with colored nodes representing different categories. (m-r) TEM images (m, o, q) and corresponding reconstructed 3D models (n, p, r), with nodes colored according to their categories, for three additional particles (QCs 2-4). Scale bars = 50 nm.

The sharp diffraction peaks observed in the 3D FFT of the tomogram exhibit 12-fold symmetry, characteristic of the dodecagonal quasicrystal (Figure 1c inset, Supplementary Movie S2). The side projection of the tomogram (Figure 1d) is also consistent with the side-view TEM image (Supplementary Figure S3), whose ordered contrast reveals a hallmark of a dodecagonal quasicrystal, namely periodic stacking of layers along the 12-fold axis and quasiperiodicity in two dimensions. Projecting the tomogram's intensities onto the $z$-axis reveals a distinct periodicity, suggesting that the particle may comprise a layered structure with alternating high and low node concentrations (Figure 1d).

Based on this periodicity, the tomogram was partitioned into 12 slices according to the mean interval between the 12 peak positions (e.g. 6.9 nm for QC-1, Supplementary Figure S4). Each slice was then projected along the $z$-axis (Supplementary Figure S5), wherein the 2D coordinates of nodes were extracted as in-plane local maxima. The $z$ coordinate of each node was separately determined by tracing the maximum cross-section of the cavity volume within the slice. Consequently, we determined the 3D nodal positions and sizes throughout the entire particle, assuming a spherical geometry for each node (Figure 1e and 1f, Supplementary Table S2).

The determined node positions and sizes were translated into a continuous 3D model, from which TEM images of the particle were simulated.[28] The simulated images agree very well with the experimental images both from the front and from the side views, suggesting the reliability of our 3D reconstruction and node-tracking procedure (Figure 1g and 1h). Notably, due to mechanical limit of the goniometer, the reconstruction suffers from resolution degradation along vertical direction,[29] with the elongation ratio $e = \sqrt{\frac{\alpha+\sin(\alpha)cos(\alpha)}{\alpha-\sin(\alpha)\cos(\alpha)}}$, where $\alpha$ is the maximum tilting angle. In our experiment, the maximum tilting angle of 60° suggests the theoretical elongation factor of $e = 1.55$. This artifact influences the morphology of cavities, as schematically illustrated in Supplementary Figure S6. Accordingly, the cavity in the tomographic reconstruction becomes elliptical, with its major axis elongated along the $z$-axis. However, the center of mass for each cage is unaffected in the elongation process, ensuring the correct extraction of nodal positions using our proposed algorithm. Besides, the nodal radius is calculated from the maximum cross-section along the $xy$-plane, which is also unaffected from the elongation.

To further reveal the structural details, the extracted nodes were categorized into three types. Type I (red) nodes correspond to the vertices of all equilateral triangles and squares in the 2D tiling pattern. The remaining nodes were further classified as either Type II (yellow) or Type III (cyan) (Figure 1i and 1j) based on their local coordination with the Type I nodes. Type II and Type III nodes are distributed alternately along the $z$-axis, as shown in the slice-resolved visualization (Figure 1k). However, some intergrowth regions are also observed (Supplementary Figure S7), which will be discussed in the defect analysis section (see below). Consequently, the entire particle comprises a complex stacking of these types of nodes (Figure 1l).

To examine the generality of the observed structural features, we applied the same tomographic reconstruction and node-categorization procedure to three additional particles (QCs 2-4, Supplementary Figures S8-S10 and Tables S3-S5). All three particles exhibit the same characteristic structural features of a dodecagonal quasicrystal (Supplementary Table S1). Tiling models were constructed based on the brightest contrast features, and the corresponding TEM images and categorized nodal positions are presented in Figures 1m-1r.

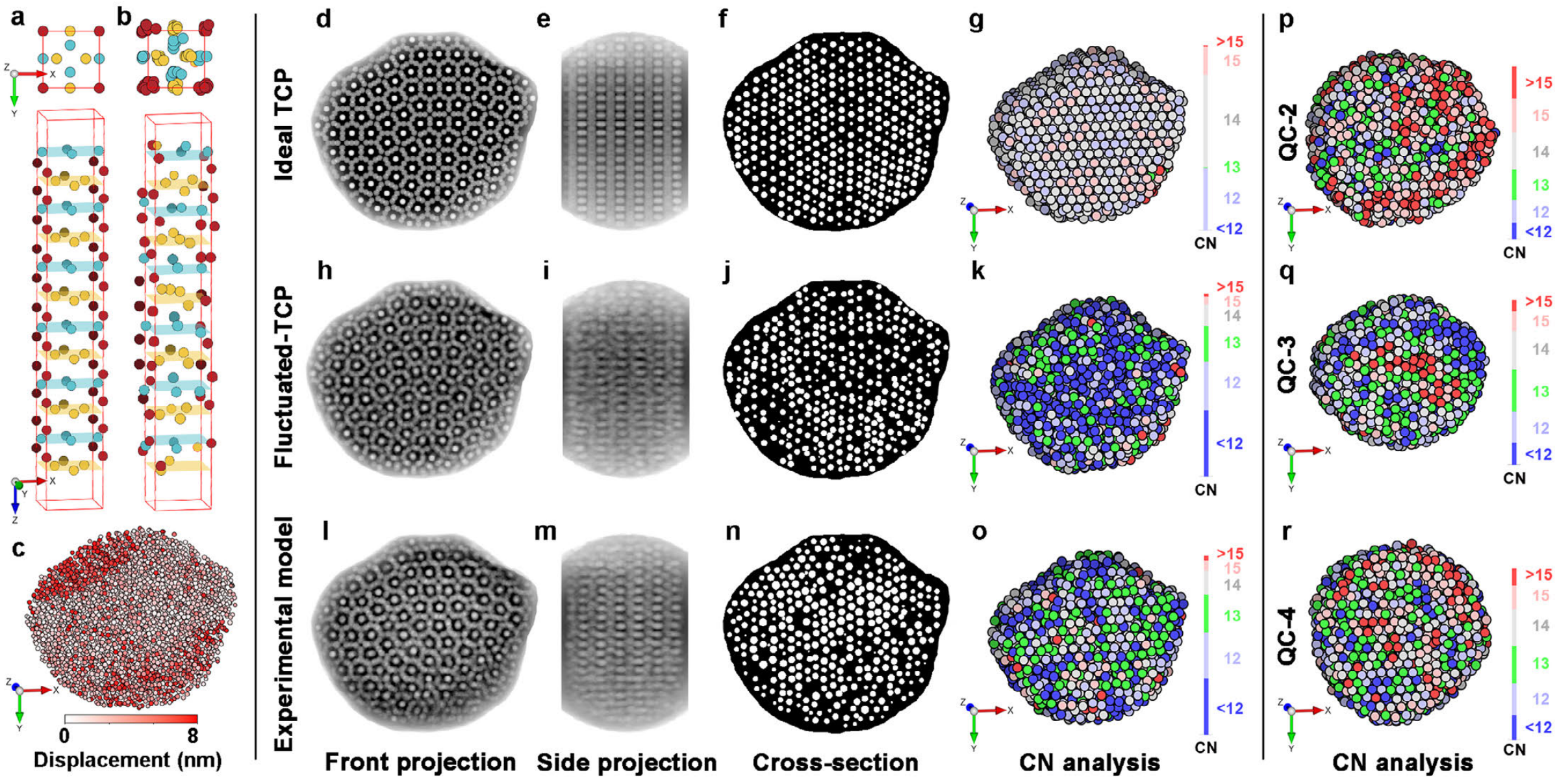

**Figure 2.** Structural analysis based on the TCP model. (a) A square-vertex unit of the ideal TCP structure, which is highly ordered with periodicity along the *z*-axis. (b) A columnar section extracted from the center of the particle (indicated by the red box in Figure 1i and 1j). The 12 colored planes correspond to the position of 12 slices shown in Figure 1d. The experimental nodal sites deviate from the ideal model. (c) Fluctuation assessment of the experimental nodal positions from the ideal TCP sites, red color indicates large fluctuation. (d,e) Front and side projections of the volumetric models of the ideal TCP. (f) Cross-section at $z = ½$ of the ideal TCP model (stack 109 of 200 along the *z*-axis). (g) Coordination number (CN) distributions of the nodes in the ideal TCP model. (h-k) The front and side projections, middle-stack cross-section, and CN analysis of the fluctuated-TCP model. (l-o) The front and side projections, middle-stack cross-section, and CN analysis of the experimental model. The ideal TCP (g) shows predominantly CN = 12, 14 and 15, whereas the fluctuated-TCP (k) and the experimental structure (o) exhibit a more evenly distributed CN numbers. (p-r) CN distributions for QCs 2-4, showing similarly broad CN distribution. The radius of the displayed sphere is enlarged by 1.4 times of the nodes for better visualization.

Previous studies have interpreted dodecagonal quasicrystals as assemblies of stacked square and triangular columns arranged quasiperiodically in the lateral direction, forming a 2D tiling. Each column is decorated with nodes in a specific manner such that the nodes are periodically arranged along the column, yielding an overall tetrahedrally close-packed (TCP) structure.[10] The Voronoi tessellation of such a TCP structure gives rise to three types of Frank-Kasper polyhedra, [$5^{12}$], [$5^{12}6^2$], and [$5^{12}6^3$], where the notation [$5^m6^n$] denotes a polyhedron with $m$ pentagonal and $n$ hexagonal faces.[30,31] In other words, an ideal dodecagonal quasicrystal can be described as a TCP structure constructed as a 3D decoration of the quasiperiodic square-triangle tiling in the 2D plane. Figure 2a illustrates a decorated square-columnar unit for the construction. Using the same classification scheme applied in Figure 1j, the Type I nodes decorating the vertices of the square in the plane occupy the $z$ = ¼ and ¾ positions of the TCP lattice, whereas the Type II and Type III nodes alternate at the $z$ = 0 and ½ positions. The 12 colored planes (cyan and yellow) in Figure 2a correspond to the $z$ centers of the 12 slices illustrated in Figure 1e. An independent structural unit therefore spans the thickness of two slices and each slice is composed of double layers (Type I+II or I+III).

To verify the TCP structure, a square-vertex unit (the red boxed region shown in Figure 1i and 1j) was extracted from the reconstructed tomogram, showing that the three node types appear in a similar alternating sequence along $z$-axis (Figure 2b). However, further subdividing the 12 slices described above into equal parts does not reveal the characteristic contrast of the expected TCP arrangement (Supplementary Figures S11 and S12); instead, the experimental nodal positions exhibit measurable local displacements. This discrepancy is further revealed by TEM image simulation.

Using the experimentally obtained square-triangle tiling and particle shapes, we constructed an ideal TCP model composed of perfect squares and equilateral triangles (Supplementary Figure S13). Since the nodes extracted from experimental data did not exhibit significant type-dependent size variations, we standardized the node dimensions to simplify the model. Notably, the ideal TCP model is highly organized, producing distinct dot-like contrast around each vertex (Supplementary Figure S14). In contrast, the experimental TEM images and the simulations based on the extracted nodal positions reveal a blurred, less ordered contrast around the vertices. Therefore, although the experimentally extracted nodal positions approximate the TCP model, local fluctuations at individual nodes may cause the actual structure to deviate markedly from the ideal TCP lattice.

To quantify these fluctuations, the displacement of each extracted node relative to its corresponding ideal TCP node was calculated. As shown in Figure 2c, most of the nodes exhibit positional fluctuations; the fluctuations in the central region are relatively small, whereas nodes located in the upper-left and lower-right corners display very large positional offsets. By adding random displacements to the node positions of the ideal TCP lattice using the calculated displacement statistics (Supplementary Table S6), a fluctuated-TCP model was constructed. The comparison of the ideal TCP, fluctrated-TCP and the experimental extracted nodes are shown in Figure 2d-2f, 2h-2j and 2l-2n. For the first two models, the elongation factor of $e = 1.55$ was also introduced for the volumetric visualizations of these models in order to simulate the missing-wedge artifact (Supplementary Figure S15). Compared with the ideal TCP model, whose nodal sites are precisely arranged in volumetric projections (Figure 2d and 2e) and whose cross-section shows an alternating pattern of large and small nodes due to different positions in successive slices (Figure 2f), the fluctuated-TCP model exhibits greater disorder. Only diffused, blur contrast is visible around each vertex (Figures 2h and 2i), and its cross-section reveals a more irregular node distribution of comparable size (Figure 2j). This fluctuated model closely matches the structure derived from experimentally extracted nodal sites (Figures 2l and 2n). Simulated TEM images (Supplementary Figure S14) and the comparasion of other three particle (Supplementary Figure S16-S18) further corroborate these observations. All these results reveal that the experimental structure exhibits a certain degree of disorder, and the fluctuated-TCP model with positional fluctuation more closely reproduces these features.

These positional fluctuations also strongly affect the coordination number (CN) of each node, which offers an objective metric for evaluating the fidelity of the proposed model to the experimental node positions. The CNs were calculated by the number of neighboring nodes with cutoff distances (1.5 times of the mean nearest-neighbor distance). To avoid bias from peripheral sites typically exhibiting low CNs, a spatial filter that excluded the outermost shell of nodes was applied. For ideal TCP, most nodes exhibit CNs of 12, 14 and 15, corresponding to $[5^{12}]$, $[5^{12}6^{2}]$, and $[5^{12}6^{3}]$ polyhedra (Figure 2g). Introducing the 3D positional shifts broadens the CN distribution, yielding a more even distribution across the ranges from below 12 to above 15 (Figure 2k). The experimentally extracted nodes in Figure 2o show a similar pattern, with a continuous distribution ranging from below 12 to above 15. The other three particles (QCs 2-4) all exhibit consistent CN statistics (Figure 2p-2r, full analysis in Supplementary Figure S19). These findings demonstrate that the fluctuated-TCP model accounts for the observed perturbation and reveal deviations between the experimental structure and the conventionally accepted ideal TCP lattice.

In TEM projections, several regions show poorly defined vertex contrast, e.g., the lower-right corner of QC-1 and the upper-right corner of QC-4. These contrasts originate from dislocations and structural distortions that arise during the self-assembly of the soft-matter system, reflecting the formation mechanism of the structure and being closely tied to its properties. However, conventional 2D TEM projections do not permit the localization of each assembly unit, which is precisely where the tomography and node-tracking approach offers a major advantage.

Figure 3a shows a mid-$z$ cross-section of QC-1 (slice 6). Although the projection of the entire particle exhibits dodecagonal quasicrystalline order, the nodal arrangement in the central region (Region-1) is topologically equivalent to that of a 2D hexagonal lattice. The upper-left part (Region-2) exhibits a similar hexagonal structure but with a different orientation, whereas the lower-right part (Region-3) appears more disordered. These distinctions are more clearly highlighted by the categorized nodes (Figure 3b), and similar partitioning is also observed in other cross-sections (Supplementary Figure S5 and S7). Consequently, the whole particle can be divided into three domains (Figure 3c and 3d). Region-2 displays similar dodecagonal quasicrystalline order to Region-1 (Figure 3c), but is rotated by 30° with respect to the $z$-axis. The boundary between the two domains is evident also from the positional fluctuations of the nodes (Figure 2c). In the side view, this region can be realized as stacking fault with layer shift along the $z$-aixs accompanied by an exchange of Type II and III nodes (Figure 3d, indicated by the cyan and yellow arrows). Meanwhile, Region-3 shows the most pronounced disorder, visible in both the front and side projections (Figure 3c and 3d). Notably, such intergrowth and disorder are not discernible in the projection of the entire particle owing to the overlap of contrast along the viewing direction (Supplementary Figure S20).

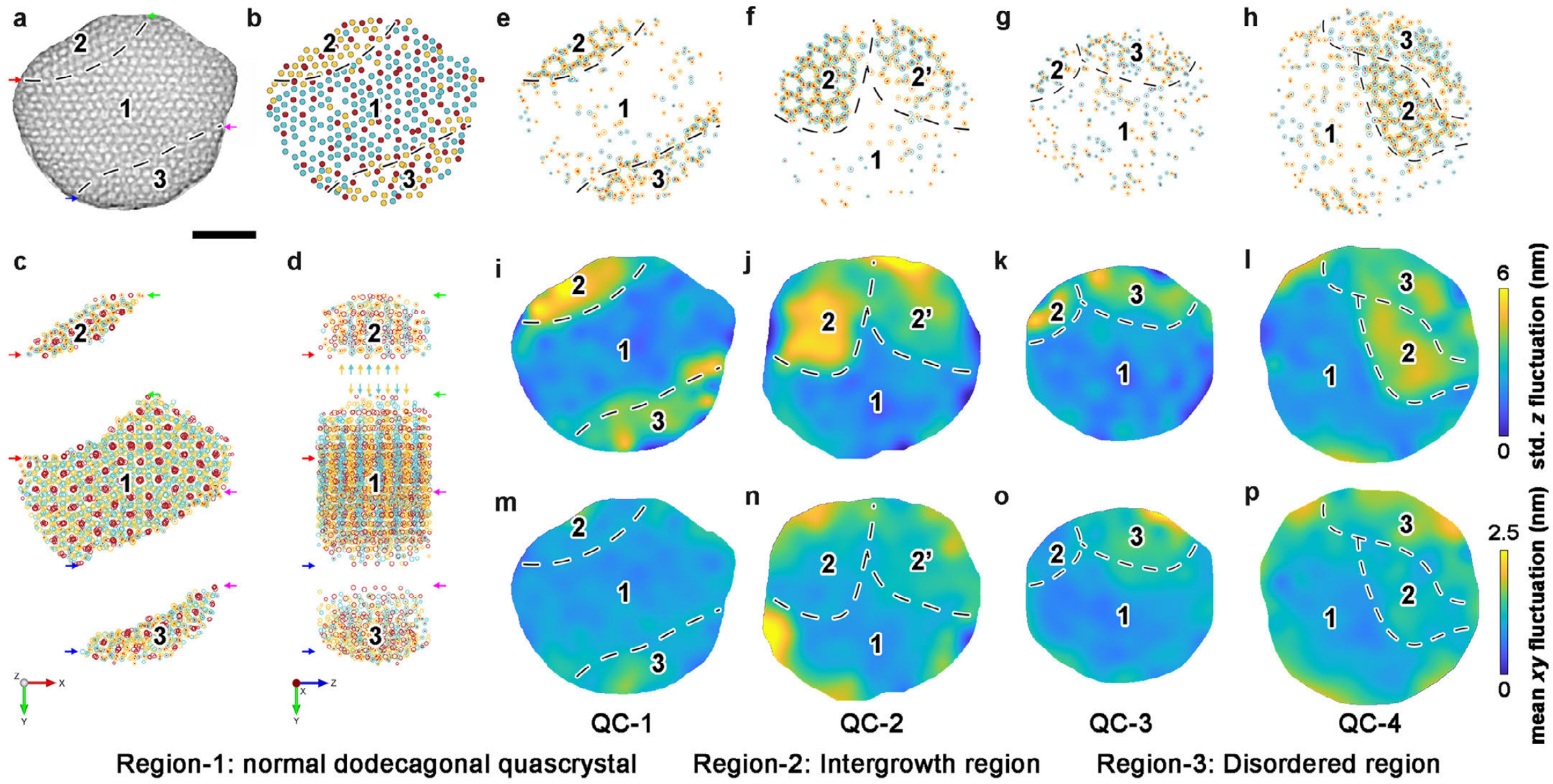

**Figure 3.** 3D analysis of defects and local disorder. (a) A cross-section of QC-1 taken at the middle position (slice 6). (b) Extracted nodes within the slice, showing that the slice structure can be partitioned into three distinct domains. Region-1 is the normal dodecagonal quasicrystal structure; Region-2 is the intergrowth region; and Region-3 is area with disordered structural arrangements. (c) Front projection of all extracted nodal positions of QC-1, separated into three domains. (d) side projection of the corresponding domains, colored arrows indicating stacking fault (intergrowth defect). (e-h) Positions of the 'outliers' with positional displacement along the *z*-axis into intermediate positions between adjacent slices in the four particles (QCs 1-4), i.e. the 'outliers', mainly concentrated in the stacking fault Region-2 and disordered Region-3. (i-l) Colormap for standard deviation of nodes positions along *z*-axis for the four particles. (m-p) Colormap for mean *xy* displacement of nodes in the four particles. Scale bar = 50 nm.

To reveal these defective nodal sites, we examined each node type separately. As shown in Supplementary Figure S21, most Type II and Type III nodes adopt a layered arrangement. However, a small portion of these nodes are displaced along the *z*-axis into intermediate positions between adjacent layers, predominantly within the misaligned and disordered regions. These displaced nodes can be termed 'outliers' (red dots in Supplementary Figure S21). From the front view in Figure 3e, these 'outliers' are predominantly located within the stacking fault region (Region-2) and the disordered region (Region-3), confirming the domain partition described as above. Similar structural features were also observed in QCs 2-4. QC-2 contains two stacking fault regions (Region 2 and Region 2′, Figure 3f), while QC-3 and QC-4 each have one stacking fault region and one disordered region (Figures 3g and 3h).

For a quantitative assessment of positional fluctuations, nodal displacements from the ideal positions were further analyzed. As the intergrowth regions are discerned by out-of-plane fluctuations, the standard deviation of the *z*-coordinate was calculated for each ideal nodal position along the periodic direction (perpendicular to *xy*-plane) and visualized as a heatmap averaged over all slices (Figure 3i-3l). The normal dodecagonal quasicrystal domains (Region-1) exhibits the smallest deviations, whereas the intergrowth domains (Region-2) shows the largest standard deviation of the *z*-coordinate; the disordered regions (Region-3) displays intermediate values. In contrast, in-plane fluctuation were quatified by computing the mean displacement of the *xy*-coordinates of nodes (Figure 3m-3p). Only the disordered domains (Region-3) demonstrates noticeable higher in-plane fluctuations. Consistent observations across all four particles confirm that the intergrowth regions have large *z*-fluctuation but low *xy*-fluctuation, while the disorder regions exhibit moderate fluctuations in both directions. The calculation for each nodal types are presented in Supplementary Figures S22-S25.

Additionally, other defect types were also identified within the structures. For example, the lower-left corner of QC-1 shows bright contrast adjacent to the regular vertices (Supplementary Figure S26a), indicating a pronounced ambiguity in pinpointing the tiling vertices. This occurs because vertex positions are generally determined by vertically aligned Type I nodes, while Type II and Type III nodes arranged in an alternating pattern. However, when these nodes are displaced, Type II or Type III nodes may overlap, thereby creating new vertices (Supplementary Figure S26b). This complicates their identification in 2D projections. Moreover, the upper-right corner of the QC-3 particle displays only a disordered contrast when main part of the structure is aligned to the electron beam; however, a modest ~3° tilt reveals the underlying tiling pattern of the upper-right part, reflecting that different regions of the QC-3 sample grew with slightly misaligned orientations (Supplementary Figure S27).

As discussed above, actual nodes deviate from their ideal TCP positions and the cross-section of the structure reveals a 2D hexagonal-like arrangement (Figure 3a and Supplementary Figure S5). The extracted nodal positions further corroborate this phenomenon: Type I nodes exhibit a pronounced shift along the *z*-axis and sit very close to the layers of Type II or III (Supplementary Table S2). Notably, the pronounced displacement is not attributed to the missing-wedge effect, which would produce histograms symmetric about zero; instead, all particles studied reveal a clear preferred orientation (Supplementary Figure S28). These observations suggest a simplified two-layer stacking model for forming the dodecagonal quasicrystalline structure. Unlike the twisted bilayer graphene quasicrystal generated by a 30° rotation with a scaling factor of √(2 + √3) in the Stampfli tiles,[5] this new bilayer model reproduces the same scaling hierarchy by relocating the nodes of an ideal 2D hexagonal lattice into specific positions.

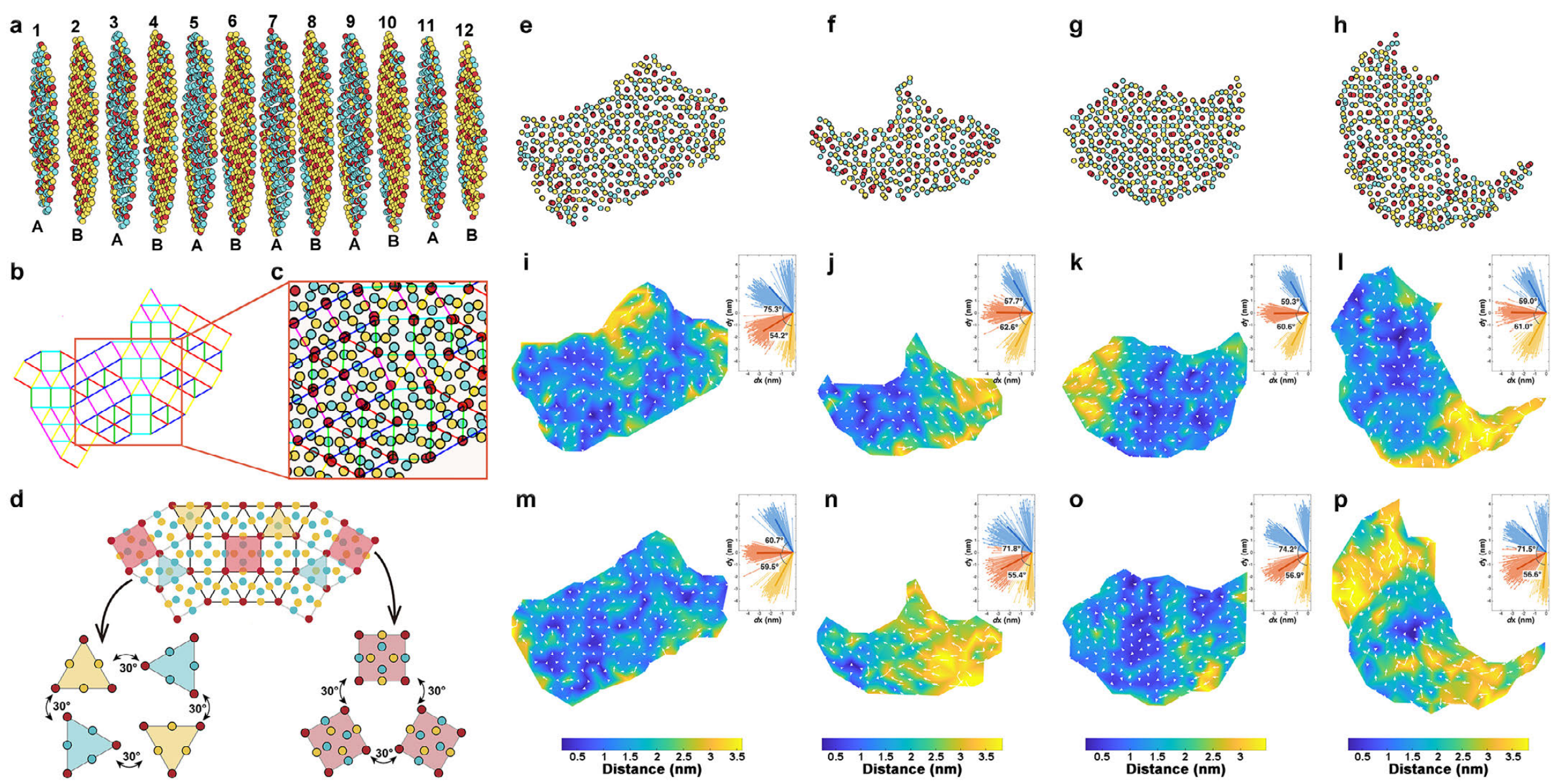

**Figure 4.** Tiling analysis and a simplified two-layer stacking model. (a) The merged 12 layers of QC-1. Odd-numbered layers (labelled layer A) consist mainly of Type I + Type III nodes, whereas even-numbered layers (layer B) are dominated by Type I + Type II nodes. (b) Dodecagonal ideal tiling of QC-1 with the intergrowth and disordered regions removed. (c) Enlarged view of the boxed region in (b), showing the nodal positions with actual tiling. (d) Schematic drawing of an ideal dodecagonal tiling. A hexagonal lattice can be obtained by rotating the basic tiling units by 30°, which creates adjacent layers. In the lower-left panel (triangular motif), the yellow background reveals a hexagonal lattice made of Type I (red) + Type II (yellow) sites (layer B), while the cyan background indicates a hexagonal lattice composed of Type I (red) + Type III (cyan) sites (layer A). In the lower-right panel (square motif), the red background displays two separate hexagonal lattices, one consisting of Type I (red) + Type II (yellow) sites and the other of Type I (red) + Type III (cyan) sites, respectively. (e-h) Overlap of the two central layers (layers 5 and 6) from four particles (QCs 1-4). The superposition of two hexagon-like layers produces the square-triangle tiling. (i-l) Colour-map of the deviation of nodes from the standard 2D hexagonal lattice in layer 5 for the four particles. (m-p) Colour-map of the deviation of nodes from the standard 2D hexagonal lattice in layer 6 for QCs 1-4. Each inset histogram in the upper-right corner of every panel shows the distance and direction of each node relative to its nearest node within a cutoff distances of 1.5 times of the mean nearest-neighbor distance.

Because the particles contain intergrowth and disordered regions, we excluded these portions and focused on the normal dodecagonal quasicrystal domain (Figure 4b). Two typical tilings, namely, $3^3.4^2$ and $3^2.4.3.4$, can be recognized and they are essential to the formation of dodecagonal quasicrystal, arising from the superposition of nodes in just two adjacent layers (layers 5 and 6). In particular, Type I nodes merge with the Type II layer at $z = 0$ or with the Type III layer at $z = ½$ of the TCP unit cell, respectively. Consequently, the particle can be described by twelve planar layers: layer A are dominated by Type I + Type III nodes and layer B mainly composed of Type I + Type II nodes, and there is a 30° rotation between two type of layers. Each layer displays a 2D hexagonal-like lattice with substantial positional displacements, and the overall structure follows an $\ldots ABAB\ldots$ stacking sequence (Figures 3a, Supplementary Figures S5 and S7). Figure 4d further demonstrates this organization from an ideal fan-shaped tiling containing subunits (triangles and squares) in different orientations, which can be extracted as rotation models every 30° and form a standard 2D hexagonal lattice each. Type I nodes align vertically to form the vertices of triangles and squares, while Type II and Type III nodes are arranged in an alternating sequence. Superimposing the two-layer structure reproduces the square-triangle tiling, a pattern that is consistent across several particles (Figure 4e-4h). It is noteworthy that triangles and squares share the same edge length (unit length), but owing to their different shapes the interplanar spacing (analogous to that of a 2D hexagonal lattice) changes from $\sqrt{3}/4$ for triangles to ½ for squares.

Accordingly, we evaluate how these nodes are shifted from ideal 2D hexagonal lattice to generate the overlapping 12-fold quasicrystal. The unit cell of the ideal hexagonal lattice was obtained by performing FFT to the central layer 6 of each particle. Due to the aforementioned 30° rotational arrangement of two type of layers, a two-layer model with 2D hexagonal-like lattice was constructed. Then iterative refinement by translating the whole lattice was carried out to minimize the total in-plane offset relative to the corresponding experimental nodal positions. Then, we calculated the displacement vectors, both direction and magnitude with each layer, of the nodes relative to a reference ideal two-layer hexagonal lattice for the four particles examined. The resulting color maps (Figure 4i-4p) display arrows indicating displacement direction, while the color scale encodes displacement magnitude, with blue representing short distances and yellow indicating longer distances. Displacements are evident in every slice and typically form localized patterns that correspond to the square and triangular motifs seen in the tiling. Moreover, the magnitude alternates between layers, every other layer shows a larger overall shift, whereas the intervening layers exhibit smaller displacements. Statistical analysis of the nearest node vectors (Figure 4i-4p insets) reveals that the angles between vectors are not random but instead close to a geometric relationship characterized by 60° of hexagonal lattice. Layer A (Type I+III) is very close to the standard hexagonal lattice, corresponding to a smaller overall displacement (Figure 4m, 4j-4l). While in layer B (Type I+II), a considerable portion of the vectors are discrete, causing the angles deviate significantly from 60°, which explains its larger overall displacement (Figure 4i, 4n-4p). Four QC paticles exhibit same pattern, the displacement of each layer of them are shown in Supplementary Figure S29 to S32 representively. These findings demonstrate that a dodecagonal quasicrystal can be realized by appropriately displacing the 2D hexagonal lattice within a two-layer structure.

## Discussion

The structure of quasicrystals remains far from being fully understood. Although higher-dimensional (nD) approaches have been developed to describe quasicrystals as 3D cuts of periodic structures in higher-dimensional space, such solutions represent average structural models that fit diffraction data and may not accurately reflect the true real-space geometry and chemistry.[11] Additionally, even highly disordered structure are presented by an averaged model with strict quasiperiodicity. Direct imaging of the local structure is essential for validating these models, yet reconstructing 3D information from 2D projections remains difficult. Atomic electron tomography (AET) shows the capability to determine disordered atomic coordinates,[32] it has not yet solved quasicrystals, partly because certain quasicrystals are beam sensitive and producing sufficiently small nanoparticles suitable for AET is challenging.

In this work, we combined electron tomography with a node-tracking algorithm to resolve the 3D mesostructure of dodecagonal quasicrystalline particles comprising ~4,000 nodes. In self-assembly systems, the constituent units are not as rigidly positioned as in atomic crystals, leading to a lower degree of structural order and enhanced structural fluctuations. Therefore, electron microscopy imaging is the only technique capable of revealing these details. The particle sizes were chosen to optimise tomography. They contain enough structural units for analysis and provide sufficient resolution to reveal detailed features and nodal sites, which was essential for a successful structural determination.

Like most soft-matter quasicrystals, our sample follows a square-triangle dodecagonal tiling. The TCP model with Frank-Kasper polyhedra has long been considered a structurasl approximant due to its topological similarities,[33,34] and the mesostructured quasicrystals were well explained by this model.[10] An atomic force microscopy (AFM) study of dendron-based dodecagonal quasicrystals suggested a similar structure based on alternating sparse-dense-sparse-dense layer stacking along the *z*-axis.[35] This packing corresponds to the Type I-II-I-III (…*ABAB*…) stacking in our model. However, the overall structure cannot be fully resolved by surface imaging alone. Significant differences in the 3D nodal arrangement and CN distribution indicate that the true structure deviates substantially from the ideal TCP model.

We also observed numerous intergrowths and defects. Although several defect models have been proposed,[36] the actual forms of these defects in 3D remain unclear. While the projected 2D TEM images of all particles display the characteristic dodecagonal tiling, the local structure is far more complex than conventional expectations. Very recently, we found that the mesostructured quasicrystal was formed by a kinetically transition, lamellar structures first appear, transition centers that incorporate tiling units then form, and these grow progressively into long-range quasiperiodicity.[37] These transition centers may arise simultaneously at different positions within a particle, giving rise to domains with distinct orientations. Defects may result from kinetic factors, with condensation of inorganic precursors stabilizing imperfect micelle arrangements.

Additionally, Type I nodes often shift vertically in columnar vertices, producing many defects relative to the TCP model. Considering the dislocations observed in atomic quasicrystals,[38] the vertical shifting of nodes in our data may imply a cross-scale structural correlation. These shifts also lead to a simplified two-layer stacking model. Overlapping alternating layers with a 30° relationship yield a superstructure with larger periodicity, only the incorporation of lattice fluctuations give rise to true quasiperiodicity. This further highlights the importance of positional fluctuations of the nodes. Notably, the vertices of the triangles and squares (Type I nodes) align vertically across layers, occupying the same positions in the *xy*-plane. This arrangement can be easily interpreted by the TCP structure, where Type I nodes are naturally separated by Type II and Type III layers. In contrast, for hexagonal-like packing, this vertical alignment does not follow standard close-packing principles. However, tomographic analysis shows that these vertices position may form continuous channels that extend vertically through the entire particle (Supplementary Figure S33). The presence of these channels indicates that micelles deformed during assembly, which changed the overall node connectivity.

## Conclusions

In summary, we have developed a methodology for fully determining and analyzing the mesostructure of dodecagonal quasicrystalline particles. By extracting the geometric centers of spherical nodes from tomographic reconstructions, this approach provides direct experimental evidence of sizes and spatial arrangements, revealing the formation mechanisms of quasicrystalline order and defects that cannot be accessed by conventional diffraction-based crystallographic methods or projected 2D images. Our results demonstrate that the complexity of quasicrystalline structures far exceeds traditional expectations, underscoring the necessity of in-depth 3D characterization. Moreover, the method is readily transferable to other materials across different length scales. When high-quality 3D tomographic data are available, the technique can universally localize structural units, grain boundaries, and defects, and perform 3D recognition and segmentation. This work offers fresh insights into the structural exploration of complex materials, including quasicrystalline, high-entropy, and amorphous systems, while broadening the analytical scope of conventional crystallography.

## Methods

**Synthesis of the dodecagonal quasicrystalline cluster.** The sample was synthesized according to the literature.[39] The gemini cationic surfactant $[C_{18}H_{37}N(CH_3)_2(CH_2)_3N(CH_3)_3]Br_2$ (denoted $C_{18-3-1}$) was used as a structure directing agent for the co-assembly with the silica precursor (tetraethyl orthosilicate, TEOS) and the organosilane (carboxyethylsilanetriol sodium salt, CES), at a molar ratio of $C_{18-3-1}$: CES: TEOS: $H_2O$ of 1: 0.6: 15: 8000.

**Characterization.** The sample was identical to the batch as reported, appropriately stored after preparation and exhibited no structural or morphological changes. Detailed characterizations are reported in Ref. 39. The TEM experiments were performed using a JEOL JEM-F200 microscope equipped with a Schottky gun operating at 200 kV (Cs 0.5 mm, Cc 1.1 mm) and a point resolution of 1.9 Å for HRTEM. Images were recorded using a GATAN OneView IS camera (4096×4096 pixels) at 50,000−120,000 times magnification under low-dose conditions.

**Electron tomography.** A tilt series of images were captured from approximately -60° to 60° with an increment of ~2°. After image denoising using the block-matching and 3D filtering (BM3D) algorithm, the alignment procedure was performed with Tomviz software,[40] which includes image shift alignment as well as axis rotation alignment. Reconstruction was performed with the REal Space Iterative REconstruction (RESIRE) algorithm.[27] The dimension of the tomogram is 400×400×400 voxels, corresponding to the resolution of ~0.48 nm. The oversampling ratio was set to 4 to retrieve the 3D real space information.

**TEM image simulation.** MesoPoreImage software[28] was adopted to simulate TEM images using obtained or generated structures based on the 3D coordinates and sizes of the nodal sites. Sample parameter: boundary roughness 0.8 nm. Scattering parameter: silica wall density 2.2 $cm^3/g$, sample thickness 100 nm, absorption factor 0.2. TEM parameter: accelerating voltage 200 kV, Cs 0.5 mm, Cc 1.1 mm, $\Delta E$ 0.3 eV, alpha 0.5 mrad, defocus -400 nm.

**CTF correction.** The CTF correction of TEM images were performed by Wiener Fliter method in Tomviz. Typical parameters are: defocus -400 nm, perfect amplitude contrast 0.2, Cs 0.5 mm, accelerating voltage 200 kV, signal-to-noise ratio 0.5.

**2D vertex tracing.** For 2D analysis, an image recognition algorithm employing template matching with geometric constraints was developed to facilitate phason-strain analysis and subsequent 3D modeling. After Gaussian filtering, the local maxima reveal a uniformly distributed set of points that includes the vertex positions and additional contrast associated with the surrounding nodal sites. Because vertex regions exhibit a characteristic bright-center, dark-surrounding pattern, template matching was used to distinguish vertices from other points. Moreover, vertices are approximately evenly spaced, allowing the position of the next vertex to be inferred from that of the preceding one. Consequently, non-vertex points can be discarded. For positions with defects and disorders, the vertex locations were confirmed manually, and the program then iterates to output the corrected vertices. The obtained 2D square-triangle tiling were subsequently used to construct the theoretical ideal TCP model by arranging perfect equilateral triangles and squares. The remaining node positions were then obtained from the centroids of triangles, the midpoints of edges between adjacent vertices, and the centroids of half-squares.

**3D node tracing.** The grayscale values of the tomogram slices were normalized using Tomviz. During node tracing process, each slice was extended along the *z*-axis by three pixels (~1.5 nm) to ensure overlap between adjacent slices and prevent omission of nodes. For each slice of cross-section, the *x*- and *y*-coordinates of nodes were extracted using a regional-maximum method. Each 2D node was then shifted along the *z*-axis to locate the *z*-coordinate where the cross-sectional area reached a maximum. The *x*- and *y*-coordinates were further refined according to the centroid of the cross-section. In a minority of cases, two local maxima maybe observed, resulting in the division of a single node into two. After all nodes were traced, those that were too close were de-duplicated based on their peak intensities (cross-sectional area). To further avoid duplicate sampling arising from the overlap-based approach, a scanning strategy along the *z*-axis with a full-slice window and a half-slice step was applied. The central half of the window was kept for all interior positions, whereas the entire window was preserved at the two ends. Finally, nodes with diameters smaller than 1 nm were removed, as their extremely small dimensions make them more likely to be artifacts than intrinsic structural features.

**Displacement analysis.** For each node, the displacement in the *xy*-plane relative to the nearest neighbor point corresponding to the ideal TCP position was calculated, and the standard deviation of this in-plane displacement distribution was obtained. Within each slice, the displacement of each node in the *z*-direction was likewise computed, and its standard deviation was determined.

**Construction of the fluctuated-TCP model.** Given that the *xy*-plane displacement field exhibits complex, non-random characteristics, it cannot be adequately captured by a simple normal distribution (Supplementary Figure S34).

To preserve directionality, the in-plane *dx* and *dy* displacements were therefore generated such that their means and standard deviations matched those of the experimental displacements, with the covariance between *dx* and *dy* likewise constrained to be consistent with the experimental data. Incorporating the generated displacement along the *z*-direction, the fluctuated-TCP model was constructed on the basis of the ideal TCP.

**Phason calculation.** The phason-strain was quantified by phason space analysis.[41]

**Determination of 'outliers'.** The determination of 'outliers' was performed exclusively for points of types II and III, with the outermost layer excluded. Nodes whose deviation from their nearest neighbor exceeded 4 nm were classified as outliers.

**Coordination number analysis.** Based on the 3D coordiates of the nodes obtained above, coordination number of each node was calculated by counting the neighboring nodes within a cutoff distances of 1.5 times of the mean nearest-neighbor distance. To exclude low-coordination surface nodes, a cuboid structuring element method based on morphological erosion was used, masking edge nodes and restricting the analysis to the particle interior. The erosion cutoff was set to roughly twice the slice thickness.

**Calculation of the node deviation from the standard 2D hexagonal lattice.** The lattice parameter of the intermediate layer were obtained through fast Fourier transform (FFT) of the central slice (slice 6), and used to generate a basic single-layer 2D hexagonal lattice. Each succesive layer was rotated by 30° and superimposed on this paper to form a 3D lattice. The obtained 3D lattice was iteratively matched to the extracted nodes. The overall *x*- and *y*-displacements were calculated layer by layer. Convergence was defined as an overall *xy* displacement of less than 0.005 nm, at which point the optimal lattice position was fixed. Then, the deviation displacement and angle from the two-layer hexagonal lattice to the nearest extracted nodes were calculated layer by layer for each node. The displacement vector of each node to its nearest neighbour node (including magnitude and direction) is analyzed using the *k*-means clustering method, as shown in the inset of Figure 4i-4p.

## Author contributions

Conceptualization: L.H.; Methodology: L.H., X.Z., X.W.; Investigation: X.Z., X.W.; Validation: N.F., L.H., O.T.; Visualization: X.Z., X.W.; Funding acquisition: L.H.; Supervision: L.H.; Writing – original draft: X.Z., L.H.; Writing – review & editing: X.W., N.F., O.T., L.H.

## Acknowledgements

The authors thank Prof. Peng Tan of Fudan University for valuable discussions and insightful comments. This work was supported by the National Natural Science Foundation of China (grant No. 22425303) and the Fundamental Research Funds for the Central Universities.

## Competing Interests

The authors declare no competing interest.